\documentclass{aa}  
\usepackage{graphicx}
\usepackage{txfonts}
\usepackage{natbib}
\usepackage{subfigure}
\def\axj{AX~J1845.0$-$0433}
\def\integral{\emph{INTEGRAL}}
\def\xmm{\emph{XMM-Newton}}
\def\asca{\emph{ASCA}}
\def\swift{\emph{Swift}}
\def\gray{$\gamma$-ray}
\def\cps{cts$\,\mathrm{s}^{-1}$}
\newcommand{\unit}[2]{\mathrm{#1}^{#2}}
\def\ecms{\mathrm{erg}\,\unit{cm}{-2}\,\unit{s}{-1}}
\def\es{\mathrm{erg}\,\unit{s}{-1}}
\def\Ms{\mathrm{M}_{\odot}}
\def\Myr{\Ms\,\unit{yr}{-1}}
\def\cnu{$\chi_{\nu}^{2}$}
\newcommand{\seefig}[1]{(see Fig.~\ref{#1})}

\def\nh{N_{\mathrm{H}}}
\def\Ec{E_{\mathrm{cut}}}
\def\CI{C_{\mathrm{ISGRI}}}
\newcommand\ra[3]{#1^{\mathrm{h}}#2^{\mathrm{m}}#3^{\mathrm{s}}}
\newcommand\dec[3]{#1\degr#2\arcmin#3\arcsec}
\begin{document}
   \title{\integral\ and \xmm\ observations of \axj\thanks{Based on observations with 1) \integral, an ESA project with instruments and science data centre funded by ESA member states (especially the PI countries: Denmark, France, Germany, Italy, Switzerland, Spain), Czech Republic and Poland, and with the participation of Russia and the USA and 2) \xmm, an ESA science mission with instruments and contributions directly funded by ESA Member States and NASA.}}

   \author{J.A.~Zurita Heras
          \inst{1}
          \and
          R. Walter\inst{2}
          }

   \offprints{J.A.~Zurita Heras}

   \institute{Laboratoire AIM, CEA/DSM-CNRS-Universit\'e Paris Diderot,
	IRFU/Service d'Astrophysique, 91191 Gif-sur-Yvette, France
	\email{juan-antonio.zurita-heras@cea.fr}
      \and
	\integral\ Science Data Centre, Observatoire de Gen\`eve, Universit\'e de Gen\`eve,
	Chemin d'Ecogia 16, 1290 Versoix, Switzerland
	\email{roland.walter@unige.ch}
   }

   \date{Received XXXX; accepted XXXX}

 
  \abstract
   {}
   {\axj\ is a transient high-mass X-ray binary discovered by \asca. The source displays bright and short flares observed recently with \integral. The transient behaviour and the bright and short flares are studied in order to understand the accretion mechanisms and the nature of the source.}
   {Public \integral\ data and a pointed \xmm\ observation are used to study in details the flaring and quiescent phases.}
   {\axj\ is a persistent X-ray binary with a O9.5I supergiant companion emitting at a low 0.2--100~keV luminosity of $\sim 10^{35}\ \es$ with seldom flares reaching luminosities of $10^{36}\ \es$. The most-accurate X-ray position is R.A.~(2000)~$=\ra{18}{45}{01.4}$ and Dec.~$=\dec{-04}{33}{57.7}$ (2$\arcsec$). Variability factors of 50 are observed on time scale as short as hundreds of seconds. The broad-band high-energy spectrum is typical of wind-fed accreting pulsars with an intrinsic absorption of $\nh=(2.6\pm0.2)\times 10^{22}\ \unit{cm}{-2}$, a hard continuum of $\Gamma=(0.7-0.9)\pm0.1$ and a high-energy cutoff at $\Ec=16_{-3}^{+5}$~keV. An excess at low energies is also observed fitted with a black body with a temperature of $kT=0.18\pm0.05$~keV. Optically-thin and highly-ionised iron (Fe~XVIII$-$XIX) located near the supergiant star is detected during the quiescence phase. The spectral shape of the X-ray continuum is constant. The flare characteristics in contrast to the persistent quiescent emission suggest that clumps of mass $M\sim 10^{22}$~g are formed within the stellar wind of the supergiant companion.}
   {}


   \titlerunning{AX J1845.0$-$0433}
   \maketitle
%

\section{Introduction}

In classical supergiant HMXB (SGXB), the persistent X-ray emission is explained by the accretion of part of the stellar wind of the supergiant companion star onto the compact object situated in a close circular orbit \citep[recent reviews of][]{Whiteal95,Psaltis06}. The observed variability has been explained in terms of wind instabilities provoked by the interaction of the compact object with the strong stellar wind \citep{Blondinal90,Blondin94}. However, transient SGXB with intensity dynamic ranges of $10^{3-4}$ have been recently discovered \citep[e.g.][]{Gonzalezal04,Smithal06}. These new sources with very low quiescent luminosities challenge the classical view of SGXB. These huge flares are likely due to the clumpiness of the strong stellar wind \citep{Zand05,Walteral07,Negueruelaal08} that is expected in OB supergiant stars \citep[][and references therein]{Oskinovaal07}. The presence of a very strong magnetic field may play an important role in the sharp intensity transition \citep{Bozzoal08}.

\axj\ was discovered with \asca\ during a monitoring of the Scutum Arm region conducted in  Oct.~18--24, 1993 (6 pointings of 20~ks each) \citep{Yamauchial95}. The source was first detected at a very low intensity for several hours, and then entered in a flaring state where the source average flux increased by a factor 100 and showed several peaks lasting tens of minutes. All detections are reported in Table~\ref{sumaxj}. The flare spectrum was fitted with an absorbed flat power law with $\Gamma=1.00\pm0.07$ and $\nh=(3.6\pm0.3)\,10^{22}\ \unit{cm}{-2}$, typical for accreting pulsars \citep{Whiteal95}. The X-ray spectrum showed no evolution. No coherent pulsation was observed with \asca\ for periods between 125~ms and 4096~s. \citet{Coeal96} pinpointed the optical counterpart as being a O9.5I supergiant star located at R.A.~(2000)~$=\ra{18}{45}{01.5}$ and Dec.~$=\dec{-04}{33}{55.5}$ (unc. $\sim1\arcsec$) and estimated a distance to the source of $\sim 3.6$~kpc. As the source is a transient supergiant high-mass X-ray binary (HMXB), \citet{Negueruelaal06} considered it as a supergiant fast X-ray transient (SFXT) candidate.

New detections of this transient were reported when \citet{Halpernal06} identified IGR~J18450$-$0435 as the hard X-ray counterpart of \axj\ after refining the \asca\ position. \citet{Molkoval04} and \citet{Sgueraal07} reported the average and flaring intensity activity of the source as observed with \integral\ and \swift\ (see Table~\ref{sumaxj}). The flare duration is typically of a few tens of minutes with sharp-rising/decaying times of a few minutes. The \swift\ X-ray spectra could be represented with similar absorbed flat power laws. Due to its recurrent fast X-ray flaring behaviour, they concur with its classification as a SFXT.

\begin{table}
\caption{Summary of previous detections of \axj.}
\label{sumaxj}      
\centering          
\begin{tabular}{c c c}
\hline\hline       
Date & Energy range & Flux \\
UTC & keV & $10^{-10}\,\ecms$\\
\hline                    
Oct. 19, 1993 13--21~h$^1$ & 0.7--10 & 0.03\\
Oct. 19 22h--20 05h, 1993$^1$ & 0.7--10 & 3\\
Mar.--May 2003$^2$ & 18--60 & 0.2\\
Apr. 28, 2005$^3$ & 20--40 & 4.5\\
Apr. 20, 2006$^3$ & 20--40 & 6.0\\
Nov. 11, 2005$^4$ & 0.2--10 & 1.1\\
Mar. 5, 2006$^4$ & 0.2--10 & 2.3\\
\hline
\end{tabular}
\begin{list}{}{}
\item[$^1$] \asca, \citet{Yamauchial95}
\item[$^2$] \integral, exposure 830~ks, \citet{Molkoval04}
\item[$^3$] \integral, peak flux, \citet{Sgueraal07}
\item[$^4$] \swift, peak flux, \citet{Sgueraal07}
\end{list}
\end{table}

In this paper, we present results from a pointed \xmm\ observation of \axj\ (PI R.~Walter) and \integral\ archival data. We present the observations and describe the data reduction in Sect.~\ref{secObs}. In Sect.~\ref{secRes}, we present the results of the refined X-ray astrometry and the temporal and spectral analysis. We discuss the temporal behaviour, the accretion and the nature of the source in Sect.~\ref{secDis}. Conclusions are summarised in Sect.~\ref{secCon}.

\section{Observations and data analysis}\label{secObs}

\subsection{\integral}

The INTErnational Gamma-Ray Astrophysics Laboratory (\integral) is a spacecraft operating since October 2002 on a 3-days highly-eccentric orbit \citep{Winkleral03} . The scientific payload is composed of four instruments. However, only data from the hard X-ray and soft \gray\ coded-mask imager IBIS/ISGRI \citep[15--300~keV, field of view (FOV) of 29$\degr$ square, angular resolution of 12$\arcmin$,][]{Ubertinial03,Lebrunal03} are going to be considered in this work.

 All public data available until march 2007 ({\it i.e.} 2075 pointings  with the source off-axis angle $<14\degr$ and exposure $>600$~s) has been analysed. These pointings are distributed on 96 spacecraft revolutions between March~19, 2003 (revolution 49, MJD~52708.4) to Nov.~21, 2005 (revolution 379, MJD~53695.5). The total exposure time on the source is 4.8~Ms spanned over 2.5~years of observations. They are not equally distributed along this period due to scheduling reasons. 

The ISGRI data are reduced using the Offline Scientific Analysis (OSA\footnote{OSA is available at {\tt http://isdc.unige.ch/?Soft+download}}) software version 6.0 that is publicly released by the \integral\ Science Data Centre (ISDC) \citep{Courvoisieral03}. Individual sky images for each pointing have been produced in the energy band 22--50~keV with typical exposure of $\sim2-3$~ks. The source count rate is extracted with the OSA tool {\tt mosaic\_spec} (version 1.4) where the position was fixed to our \xmm\ position and the point-spread function width was fixed to 6$\arcmin$. Detections are considered at a significance $\geq 5.1\sigma$. To search for the source quiescent level, we built deep-sky mosaics (with exposures 50--1000~ks) disregarding all the pointings where the source is detected. Detection of the source in deep mosaics are considered at a significance $\geq 6\sigma$. Two methods are used to extract the spectrum of the source. If the source is significantly detected in each individual pointing, the source spectra are extracted for each pointing with {\tt ii\_spectra\_extract} and are summed to build the average spectrum using {\tt spe\_pick}. Otherwise, the spectra are extracted from the deep-mosaic images with {\tt mosaic\_spec}. The redistribution matrix file (RMF) {\tt isgr\_rmf\_grp\_0023.fits} is rebinned into 5 channels spread between 15 and 80~keV. Lightcurves with a binning of 10~s are extracted for each individual flares with {\tt ii\_light} version 8.2. The lightcurves are given in counts$\,$s$^{-1}$. The 22--50~keV count rates can be converted with the relation $1\,\mathrm{Crab}=117\,\mathrm{counts}\,\mathrm{s}^{-1}=9\times 10^{-9}\ \ecms$.

\subsection{\xmm}\label{secObsXMM}

The main scientific instrument on-board the X-ray Multi-Mirror Mission  \citep[\xmm,][]{Jansenal01} satellite is the EPIC camera composed of two MOS \citep{Turneral01} and one pn \citep{Struderal01} CCD cameras. It has imaging, timing, and spectral capabilities in the 0.2--12~keV energy range with a 30$\arcmin$ FOV. The EPIC cameras were operating in imaging science mode with full window and with medium filters.

 \axj\ was observed with \xmm\ on April 3, 2006, from 15:01:33 to 20:21:46 UTC (MJD~53828.627--53828.849) for a total exposure of 19~ks. There is no simultaneous observation with \integral.
 
Event lists for each MOS and pn cameras are generated with the Science Analysis Software (SAS\footnote{SAS is available at {\tt http://xmm.vilspa.esa.es/external\\/xmm\_sw\_cal/sas.shtml}}) version 7.0.0 using the {\tt emproc} and {\tt epproc} tool, respectively. The observation does not suffer from enhanced background features, so all events are kept for a total exposure of 19~ks for both MOS and 17~ks for pn. Images for MOS[12] and pn are generated with 2$\arcsec$ and 4$\arcsec$ resolution, respectively, using good events up to the quadruple level in the 0.8--10~keV energy range and disregarding bad pixels. An accurate X-ray position determined with EPIC is calculated with the SAS task {\tt edetect\_chain}. Four images with energy ranges 0.5--2, 2--4.5, 4.5--7.5, and 7.5--12~keV for each MOS and pn cameras are extracted. Then, the best position for each individual EPIC camera is determined. Finally, the best source position is calculated as the mean of the positions given by the three cameras. The source location accuracy is limited by the spacecraft astrometry that is of 2$\arcsec$\footnote{The \xmm\ astrometric accuracy is described in the calibration document {\tt XMM-CAL-TN-0018}.}, the statistical error of $0.04\arcsec$ being insignificant in comparison.

In the EPIC/pn image, the event list of the source$+$background is extracted from a region of 30$\arcsec$ radius centered on the source. A background event list is extracted in the same CCD at the same distance of the read-out node from a region of similar size not affected by the bright source. Spectra are extracted selecting only single events but disregarding bad pixels. Specific RMF and ancillary response files (ARF) are generated with the SAS tasks {\tt rmfgen} and {\tt arfgen}, respectively. The {\tt Xspec} version 11.3.2t package \citep{Arnaud96} is used to plot and fit the resulting spectra corrected from background. For light curves, single and double events are selected. The source light curve is corrected from the background using the SAS task {\tt lccorr}. We applied the barycentric correction with the SAS task {\tt barycen}.

\section{Results}\label{secRes}

\subsection{Refined X-ray position}

A 22--50~keV ISGRI image has been generated from pointings where the source is detected \seefig{isgri_lc}. The source is detected at a level of 17$\sigma$ for a total exposure of 17.3~ks.
The best  hard X-ray position is R.A.~(2000)~$=\ra{18}{45}{07.8}$ and Dec.~$=\dec{-04}{33}{01.2}$ with an error of 1.8$\arcmin$ \citep{Grosal03}. This position is consistent with the position given in the 3rd ISGRI catalogue \citep{Birdal07}.

\begin{figure}
\centering
\resizebox{\hsize}{!}{\includegraphics{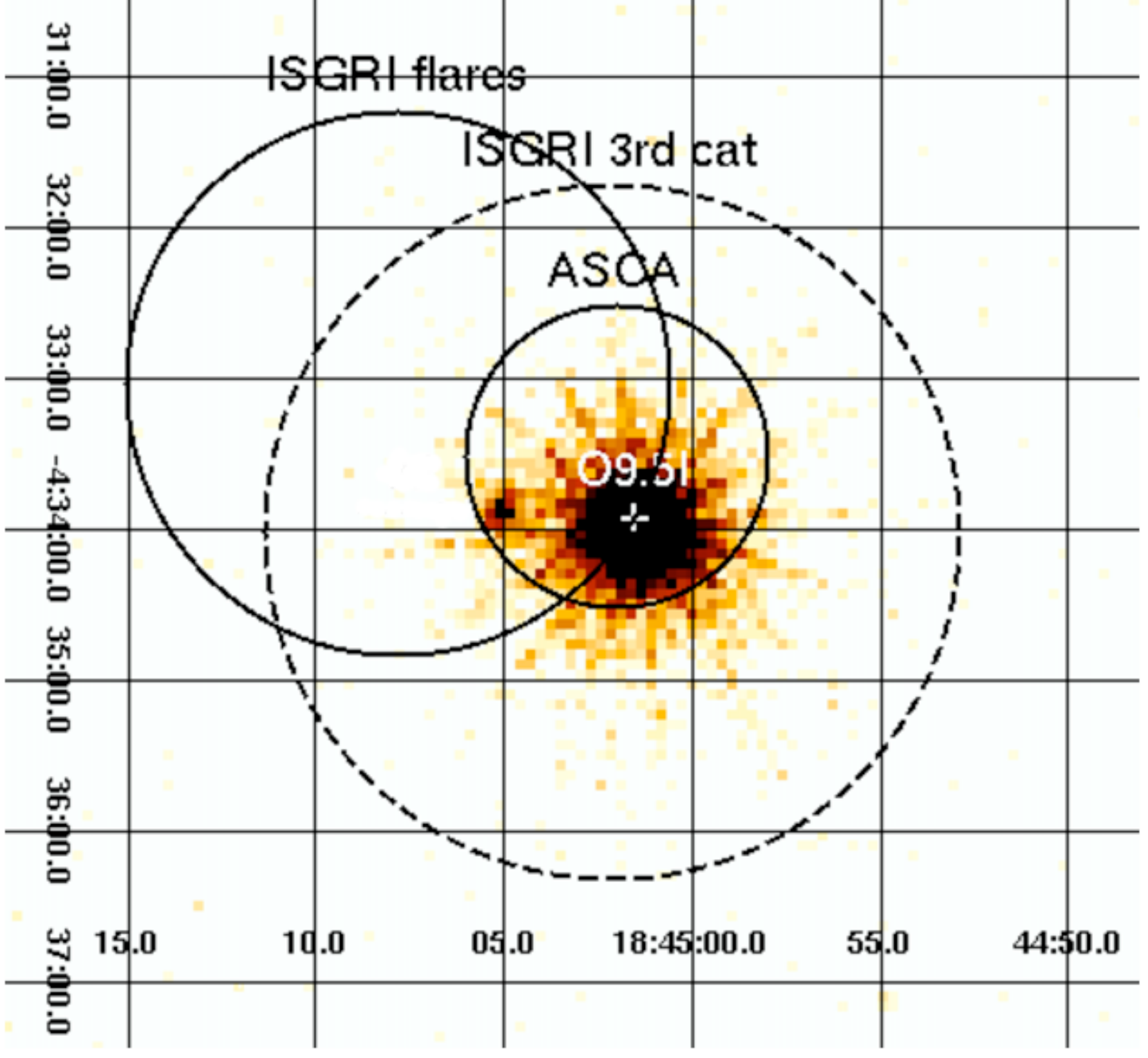}}
\caption{EPIC/MOS1 image in the 0.8--10 keV energy range. The optical counterpart of \axj\ is indicated with a cross labelled with its spectral type. Its position coincides with that of the X-ray source. Only the ISGRI (derived from the flares and from the 3rd ISGRI catalogue) and the ASCA error boxes are reported for visibility. The \swift\ position is well embedded in the \xmm\ source.}
\label{mos1_image}
\end{figure}

One bright source corresponds to \axj\ position in EPIC images \seefig{mos1_image}. The best position determined with EPIC is R.A.~(2000)~$=\ra{18}{45}{01.4}$ and Dec.~$=\dec{-04}{33}{57.7}$ with an uncertainty of 2$\arcsec$. It is located 2.9$\arcsec$ away from the optical counterpart given by \citet{Coeal96}.
This X-ray refined position is consistent with the position of the counterpart within the respective error boxes. It is also compatible with the \asca\ and \integral\ positions. The \swift\ position is outside the EPIC error box (8$\arcsec$ away), however no \swift\ uncertainty is given in \citet{Sgueraal07}.

\subsection{Timing analysis}

\subsubsection{Long-term variability}

\begin{figure}
\centering
\resizebox{\hsize}{!}{\includegraphics{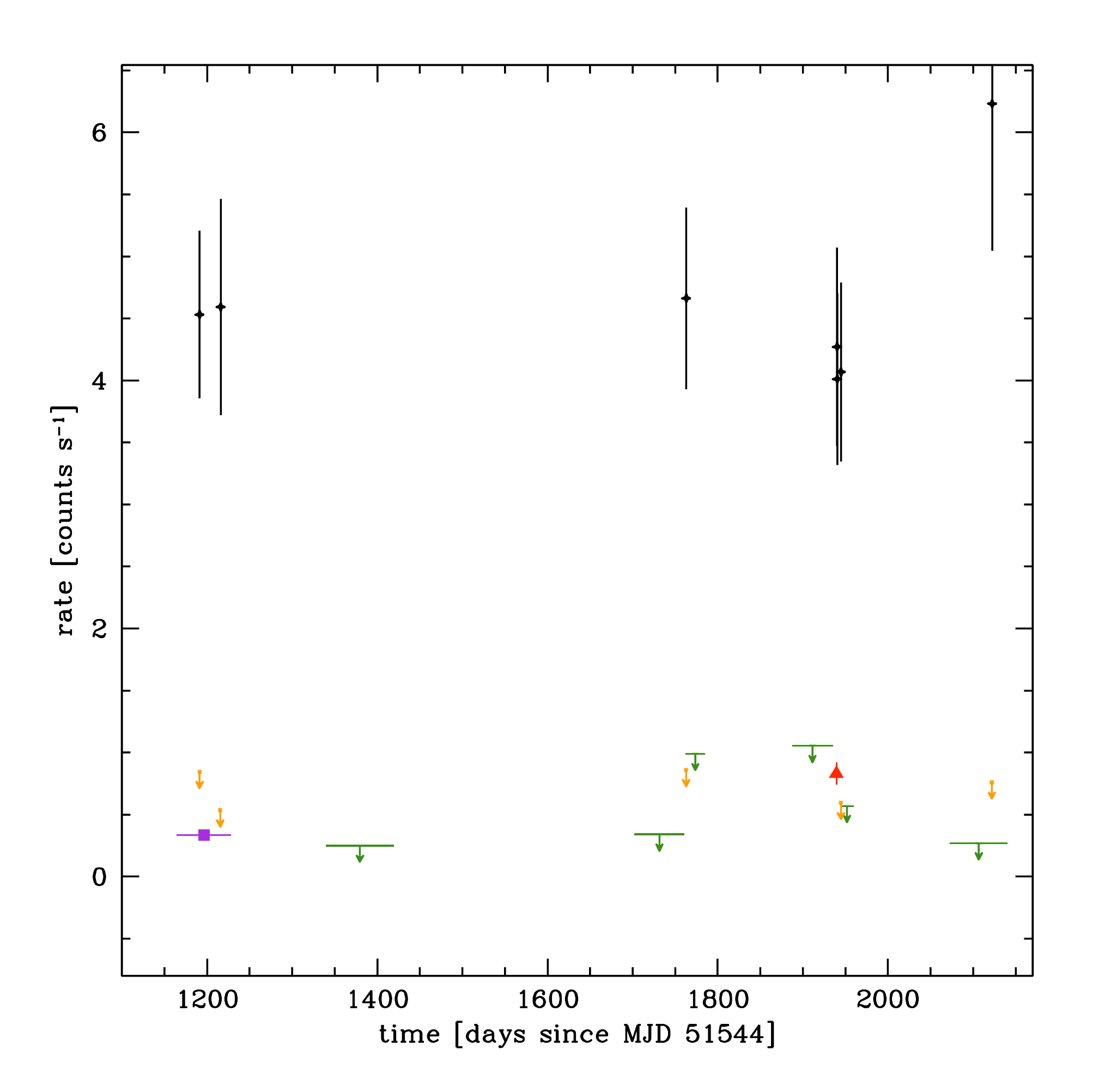}}
\caption{ISGRI 22--50~keV long-term lightcurve of \axj. Detections in single pointings with exposure $\sim2$~ks are shown with crosses and error bars. Disregarding the flares, the triangle corresponds to the average source flux during revolution {\tt 308}. The box corresponds to a detection in the first deep mosaic with an exposure of 1.2~Ms. Otherwise, 5$\sigma$ upper limits are represented (exposures of $\sim10^{2-3}$~ks).}
\label{isgri_lc}
\end{figure}

The 22--50~keV long-term variability of \axj\ is shown in Fig.~\ref{isgri_lc}. The source is only detected in 7 pointings out of 2075 with average (over the pointing duration) count rates within 4--6~\cps\ ($=$34--51~mCrab) (cross points in Fig.~\ref{isgri_lc}). These rare detections in single pointings are considered as flares. They are randomly distributed along the 2.5~yr observations. Two detections are separated by 4~h. This could be the signature of  a longer flare as the source is at the limit of detectability (see Table~\ref{tab_flare}).
\begin{table}
\caption{List of flares of \axj\ observed with \integral/IBIS/ISGRI.}             
\label{tab_flare}      
\centering          
\begin{tabular}{c c c c c c}
\hline\hline       
Time & 22--50~keV Flux & On-time exp. & Pointing\\
MJD$-$51544 & \cps & ks & ~\\
\hline                    
1191.2173 & $4.5\pm0.7$ & 2.2 & {\tt 00580043}\\
1216.0800 & $4.6\pm0.9$ & 1.8 & {\tt 00660092}\\
1763.0657 & $4.7\pm0.7$ & 3.0 & {\tt 02490072}\\
1940.2455 & $4.3\pm0.8$ & 2.0 & {\tt 03080097}\\
1940.4107 & $4.0\pm0.7$ & 3.5 & {\tt 03080103}\\
1944.8607 & $4.1\pm0.7$ & 2.0 & {\tt 03100044}\\
2122.4849 & $6.2\pm1.2$ & 3.0 & {\tt 03690058}\\
\hline                  
\end{tabular}
\end{table}

We built mosaics for each of the 96 revolutions discarding the 7 pointings reported above. The source is only detected in revolution {\tt 308}: exposure 159~ks, significance $9\sigma$, and count rate $0.83\pm0.09$~\cps\ ($=7.1\pm0.8$~mCrab) (triangle in Fig.~\ref{isgri_lc}).  In order to avoid contamination near the flares, we also tested the presence of the source by discarding the last part of the revolution stopping 20~ks before the flares. For an exposure of 134~ks, the source is still detected at a $7.4\sigma$ level for a count rate of $0.7\pm0.1$~\cps. The source varies by a factor 5 between the flares and the low-level emission during revolution {\tt 308}. For the other 5 revolutions in which at least one flare occurred ({\it i.e} {\tt 58}, {\tt 66}, {\tt 249}, {\tt 310} and {\tt 369}), the source is not detected with a $5\sigma$ upper limit in the range $0.5-1.0$~\cps\ ($=$4.5--8.5~mCrab) reported as cross-arrows in Fig.~\ref{isgri_lc}. During rev. 308, we have evidence for a long period of activity, but there is no evidence for such behaviour in the other revolutions where only one isolated flare was detected.

Data from revolutions with no detection are cumulated together over periods of 30~days. The source is only detected in the first deep mosaic of 1.2~Ms at 9$\sigma$, and with a count rate of $0.33\pm0.04$~\cps\ ($=2.8\pm0.3$~mCrab) (box in Fig.~\ref{isgri_lc}). For the other 6 mosaics, the 5$\sigma$ upper limit level lies within $0.2-1.0$~\cps\ ($=0.25-8.5$~mCrab).

Data from the 6 deep mosaics with non-detection are cumulated together to build a very-deep mosaic. The source is detected at a 9.2$\sigma$ level with $0.25\pm0.03$~\cps\ ($=2.1\pm0.3$~mCrab) for an exposure time of 2.6~Ms. This corresponds to a factor 25 lower in comparision with the brightest flare detected in pointing {\tt 03690058}. We can conclude that the quiescent level is $\sim2.5$~mCrab.

\subsubsection{Short-term variability}

\begin{figure*}[htbp]
  \centering
  \mbox{
    \subfigure[Flare state (bin=20~s).]{\includegraphics[width=0.5\textwidth]{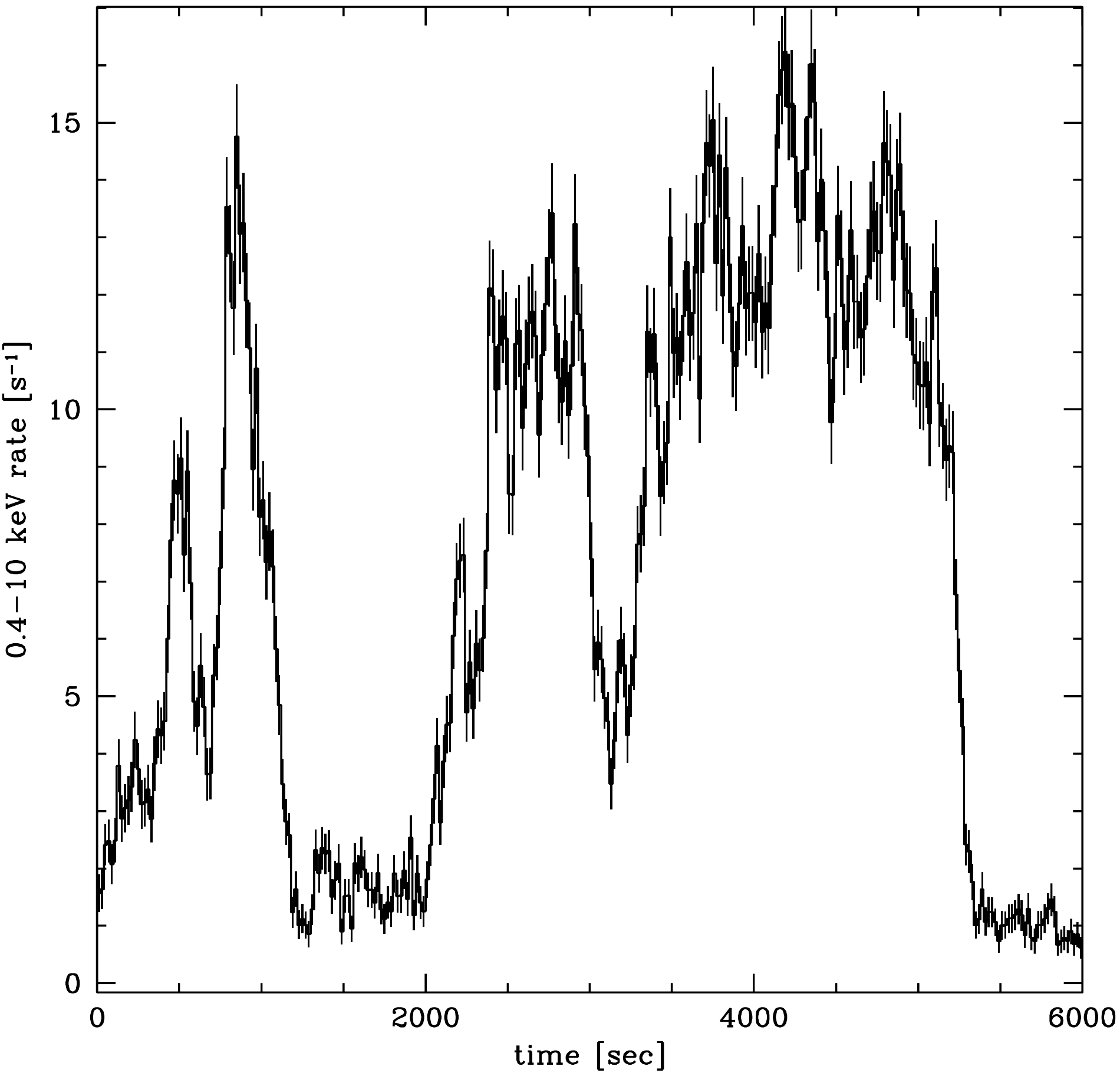}}
    \subfigure[Quiescent state (bin=50~s).]{\includegraphics[width=0.5\textwidth]{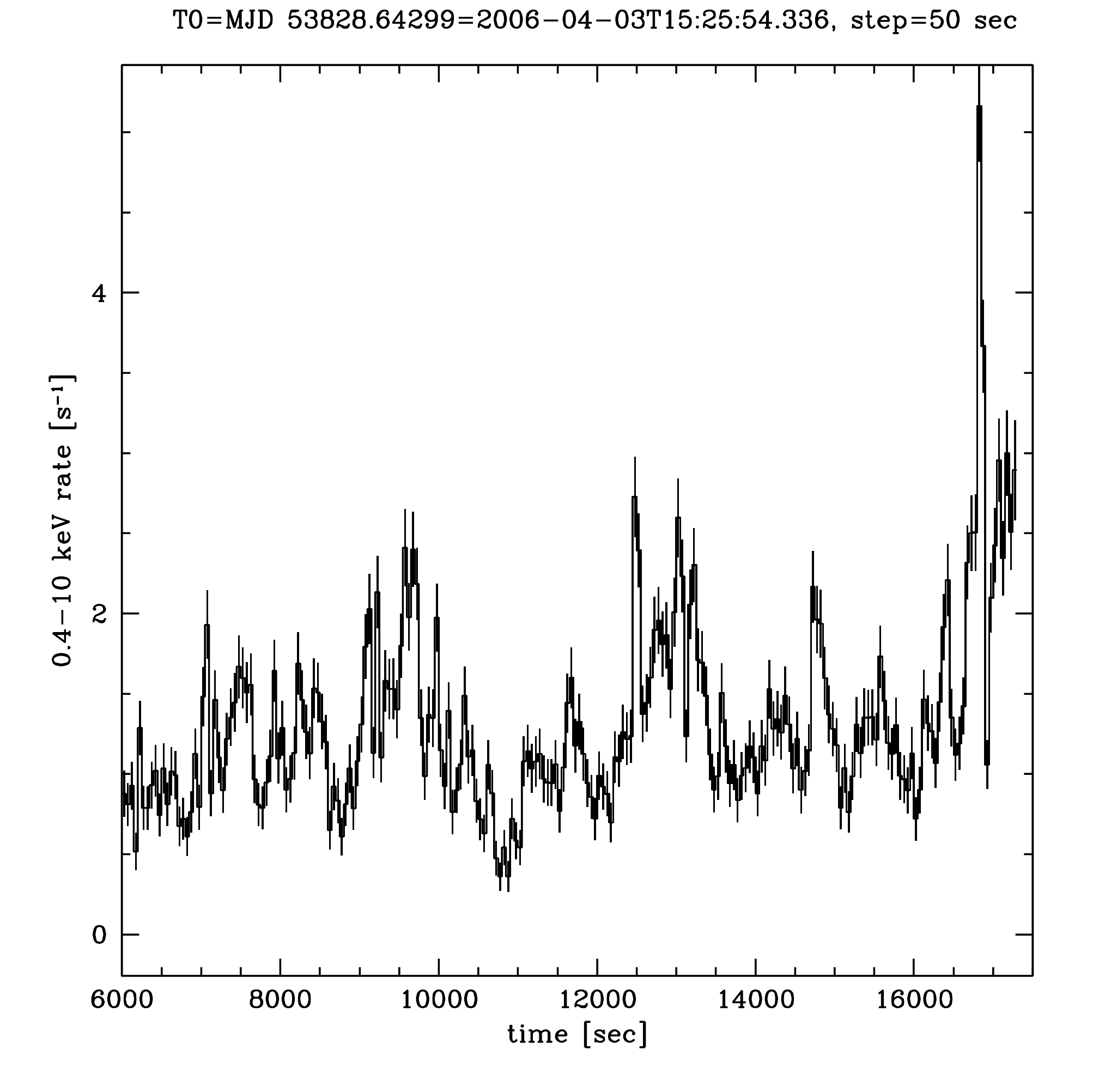}}
  }
  \caption{EPIC/pn 0.4-10~keV lightcurve of \axj\ divided in two parts to show the flare ({\it left}) and quiescent ({\it right}) state (T0=MJD 53828.6430).}
  \label{pn_lc_zoom}
\end{figure*}

The EPIC/pn 0.4--10~keV light curve is displayed in Fig.~\ref{pn_lc_zoom}. During the first 5~ks, \axj\  displays two short and complex flares lasting 1~ks and 3~ks, respectively. Both flares behave similarly with sharp rises/decays of the intensity that vary between $\sim2-12$~\cps\ on time scales of few hundred of seconds. During the flares, variations are also important with short peaks reaching $\sim 15$~\cps\ (maximum $16.2\pm0.6$~\cps). Moreover, abrupt and short ($\lesssim 300$~s) decays of the intensity down to $\sim4$~\cps\ happen within the flares but without reaching the quiescent level. The flare durations are comparable to those observed by \integral.

After the two flares, the source remains at a low level (average count rate of $1.2\pm0.2$~\cps) and shows continuous variability of a factor 2--3 with several peaks $>2$~\cps\ (see Fig.~\ref{pn_lc_zoom}(b)). Taking into account the whole observation, the variability factor between the lowest and highest 0.4--10~keV count rate is 45. Almost at the end of the observation, there is a flare-like event lasting only $\sim$50~s that reaches $5.2\pm0.3$~\cps.

We also searched for pulsation in the range 0.1--5000~s in the EPIC/pn lightcurve. However, no pulsation was observed neither during the flare nor during the quiescence phase.

\subsection{Spectral analysis}

The source shows two emission states in the X-rays: a quiescence emission randomly interrupted by rare sharp and short flares. Therefore, one X-ray spectrum is extracted for each state. Spectrum events lists are selected from good time intervals created from the 0.4--10~keV light curve using the criteria that the rate is $\geq$6~\cps\ for the flare and $\leq$3 \cps\ for the quiescent spectra, respectively. The spectral bins are grouped to have at least 100 counts and 50 counts per channel for the flare and quiescent spectra, respectively. Therefore, the $\chi^{2}$ statistic is used throughout the spectral analysis (\cnu\ is used to note the reduced $\chi^{2}$). Concerning the hard X-rays, the flare spectrum is extracted combining the seven detections. The quiescent spectrum is extracted from the deep mosaic image corresponding to the 9$\sigma$ detection. 

\begin{figure}
\centering
\resizebox{\hsize}{!}{\includegraphics{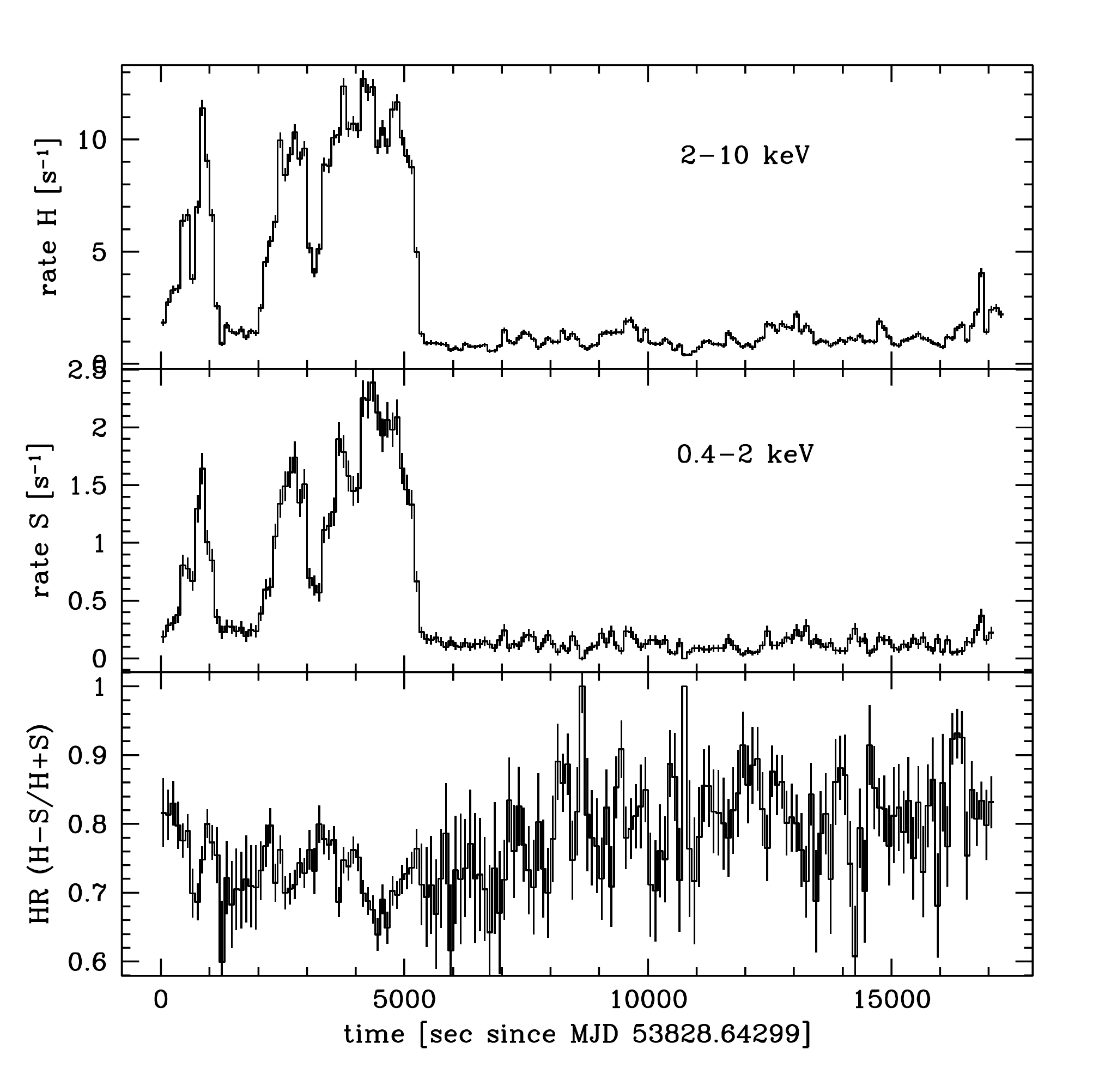}}
\caption{EPIC/pn hardness ratio lightcurve of \axj\ (bin 100~s).}
\label{pn_hr}
\end{figure}
In order to search for spectral variations, the hardness ratio ($HR\equiv(H-S)/(H+S)$) is built from two light curves with energy ranges of  $S\equiv 0.4-2$~keV and $H\equiv 2-10$~keV \seefig{pn_hr}. The HR varies between the flare and the quiescence phases with $<HR_{\mathrm{flare}}>=0.717\pm0.005$ and $<HR_{\mathrm{quiesc}}>=0.795\pm0.007$, indicating a slight softening of the spectrum during the flare.

EPIC/pn and ISGRI spectra are fitted together using a multiplicative constant ($C_{\mathrm{ISGRI}}$) to take into account the cross-calibration between instrument. The constant is fixed to 1 for EPIC/pn and remains free for ISGRI. Photoelectric absorption is included in all models. A powerlaw (PL) or a black body (BB) clearly fail to fit the flare spectrum with \cnu\ of 1.76 and 1.94 (both 207 dof), respectively. An energy break is needed at high-energy. Therefore, a broken PL and a PL with a high-energy cutoff are used (in {\tt Xspec}, {\tt bknpower} and {\tt cutoffpl}, respectively, thereafter BKN and CPL). The \cnu\ is improved with 1.49 (205 dof) and 1.18 (206 dof), respectively. A slight excess is observed at low energies ($<2$~keV). The addition of a BB with low temperature $kT\sim0.1$~keV improve the \cnu\ down to 0.95 (204 dof). 

In the case of the quiescent spectrum, absorbed PL or BB models fail to fit the data for a reasonable cross-calibration constant. Therefore, we apply the BKN and CPL models. The CPL model gives again the best fit with a \cnu\ of 1.43 (125 dof). A drop of the spectrum data is clearly visible at energies higher than 8~keV in the EPIC/pn part compared to the model. An edge multiplicative component is added to the model improving the \cnu\ down to 1.17 (121 dof).
A soft excess at energies lower than 2~keV is also observed in the quiescent spectrum, so we also add a soft BB ($kT\sim0.1$~keV). This results in a final \cnu\ of 1.00 (121 dof).

The soft component observed in both quiescent and flaring spectra can also be modelled with the emission of an optically-thin plasma \citep[{\tt mekal},][and references therein]{Kaastra92}. The results are almost identical between the {\tt bb} and {\tt mekal} models with similar \cnu\ $\sim0.9$ (324 dof).

\begin{figure}
\centering
\resizebox{\hsize}{!}{\includegraphics[angle=90]{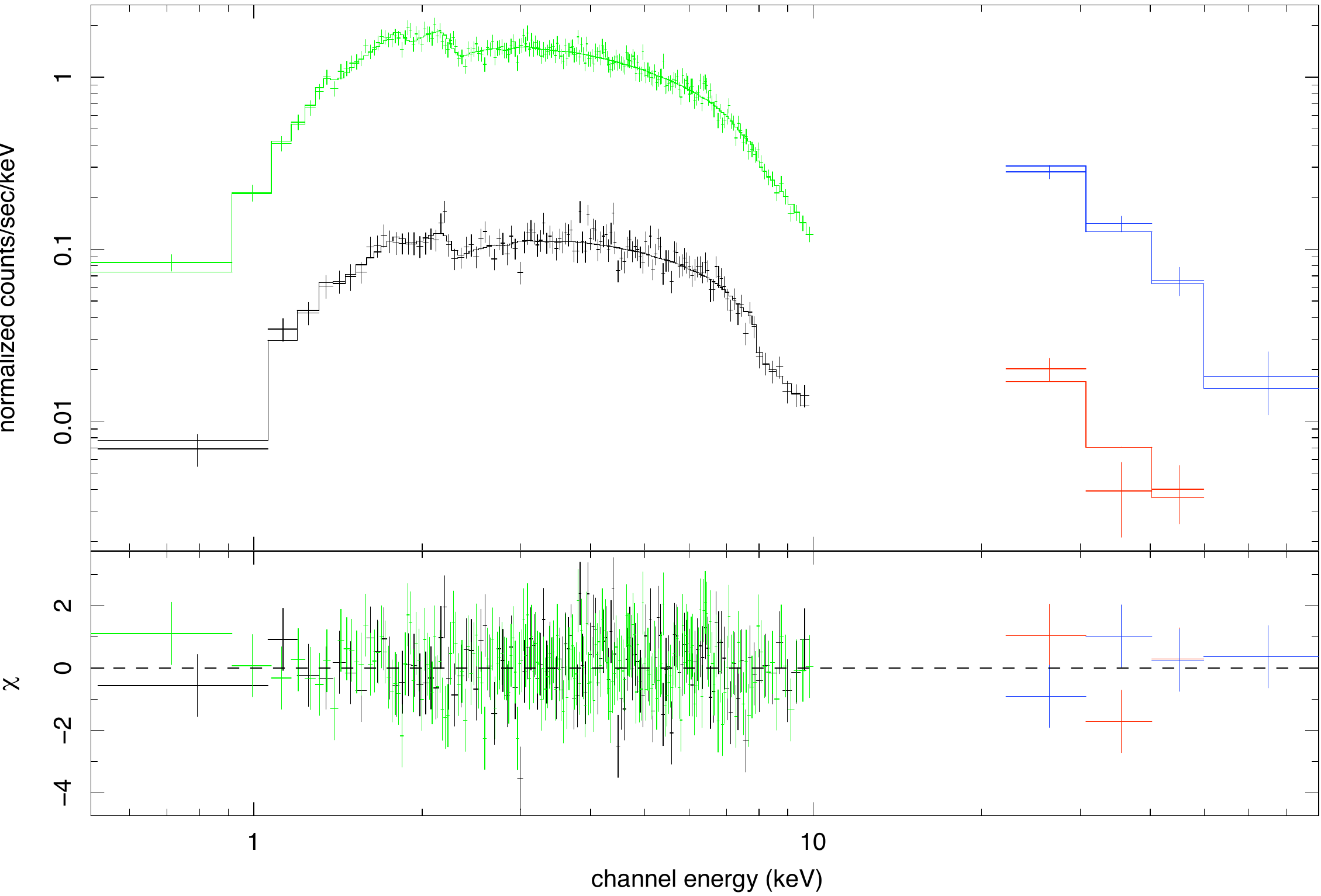}}
\caption{Broad-band spectra of \axj. The flare spectrum is displayed on top and the quiescent one at the bottom. The spectra are fitted simultaneously with the following model: {\tt cons*wabs*edge*(mekal+cutoffpl)}. Residuals are shown in the bottom panel.}
\label{ima_spec}
\end{figure}

The source spectrum is made of two components with a thermal component dominating at low energies ($\lesssim 2$~keV) and a hard power law continuum with a high-energy cutoff ($>10$~keV), to which an inter-calibration constant, the absorption, and an edge are added\footnote{The {\tt Xsepc} model is: {\tt cons*wabs*edge*(mekal+cutoffpl)}.}. The parameters are let to freely vary between both flare and quiescent spectra except the edge energy that is fixed to the value found in the quiescent spectrum and the high-energy cutoff fixed to the value found in the flaring state. The column density and the temperature of the soft component do not significantly change between the flare and quiescent spectra. Therefore, these parameters are simultaneously fitted:
\begin{itemize}
\item both spectra: $\nh=(2.6\pm0.2)\,10^{22}\ \unit{cm}{-2}$, $E_{\mathrm{edge}}=7.9\pm0.1$~keV, $kT=0.18\pm0.05$~keV and $\Ec=16_{-3}^{+5}$~keV;
\item flare spectrum: $N_{\mathrm{mekal}}=0.8_{-0.6}^{+5.7}\ \unit{cm}{-5}$, $\tau_{\mathrm{edge}}=0.19\pm0.08$, $\Gamma=0.9\pm0.1$, $\CI=2.3_{-0.5}^{+0.7}$ and the unabsorbed fluxes are $1.2\times10^{-9}\ \ecms$ (0.4--10~keV) and $2.9\times10^{-10}\ \ecms$ (20--100~keV); and
\item quiescent spectrum: $N_{\mathrm{mekal}}=0.06_{-0.05}^{+0.32}\ \unit{cm}{-5}$, $\tau_{\mathrm{edge}}=0.5\pm0.1$, $\Gamma=0.7\pm0.1$, $\CI=0.9_{-0.3}^{+0.4}$ and the unabsorbed fluxes are $0.9\times10^{-10}\ \ecms$ (0.4--10~keV) and $1.7\times10^{-11}\ \ecms$ (20--100~keV),
\end{itemize}
with $\chi_{\nu}^{2}\sim0.9$ (326 dof) \seefig{ima_spec}.

\section{Discussion}\label{secDis}

The \xmm\ astrometry confirms the supergiant counterpart of \axj\ reported by \citet{Coeal96}.

\subsection{Persistent source}

For a distance of 3.6~kpc, the quiescent average 0.4--10 and 20--100~keV luminosities of \axj\ are $1.4\times10^{35}\ \es$ and $2.7\times10^{34}\ \es$, respectively. During the flares, the average 0.4--10 and 20--50~keV luminosities are $1.8\times 10^{36}\ \es$ and $4.5\times 10^{35}\ \es$, respectively. During the quiescence phase, the source features continuous variability (of factor 2, see Fig.~\ref{pn_lc_zoom}). Discarding these mini-flares,  the minimum source 0.4--10~keV luminosity becomes as low as $1.1\times 10^{34}\ \es$. 
The emission observed with \swift/XRT \citep[see Fig.~4 and 5 in][]{Sgueraal07} is also compatible with the quiescent emission observed with EPIC. This quiescent emission has also been observed with \asca\ (still estimated with a rather poor statistic). This persistent and variable emission indicates that the source must be permanently accreting as this luminosity is too high to be explained by the X-ray luminosity of $10^{30-33}\ \es$ detected in single OB stars \citep[][and references therein]{Berghoeferal97}. 

Let considers a compact object with a mass of $M_{\mathrm{X}}=1.4\ \Ms$ and radius $R_{\mathrm{X}}=10$~km. For a O9.5I supergiant star, the likely stellar parameters are $M_{*}\sim30\ \Ms$, $R_{*}\sim23\ R_{\odot}\sim 1.6\times10^{12}$~cm, $log(L_{*}/L_{\odot})\sim5.5$ and $T_{}\sim 3\times10^{4}$~K \citep{Martinsal05}. The average parameters of the stellar wind for a O9.5I massive star are a terminal velocity of $v_{\infty}=1765\ \mathrm{km}\,\unit{s}{-1}$ \citep{Prinjaal90} and an escape velocity of $v_{\infty}/v_{\mathrm{esc}}=2.6$ \citep{Lamersal95}. These values imply a theoretical stellar mass-loss rate of $\dot{M}_{\mathrm{wind}}\sim9.4\times 10^{-7}\ \Myr$ \citep[see Eq.~12 in][]{vinkal00}. The average mass-accretion rate can be estimated as $\dot{M}_{\mathrm{accr}}=L_{\mathrm{X}}/\epsilon c^{2}=3.5\times 10^{-11}\ \Myr$ where $\epsilon=0.1$ is the accretion efficiency and $c$ the speed of light. The accretion radius is $R_{\mathrm{accr}}=2GM_{\mathrm{X}}/(v_{\mathrm{wind}}^{2}+v_{\mathrm{X}}^{2})$ where the stellar wind velocity is $v_{\mathrm{wind}}=v_{\infty} (1-R_{*}/r)^{\beta}$ with $\beta\approx 1$ \citep[][and reference therein]{Kudritzkial00} and $v_{\mathrm{X}}$ is the compact object velocity. $\dot{M}_{\mathrm{accr}}$ and $\dot{M}_{\mathrm{wind}}$ are related as $\dot{M}_{\mathrm{accr}}=(R_{\mathrm{accr}}^{2}/4a^{2})\,\dot{M}_{\mathrm{wind}}$ where $a$ is the binary separation. In order to explain a persistent quiescent luminosity of $\lesssim 2\times10^{35}\ \es$ in \axj, one needs a binary separation $a\gtrsim 5\times10^{12}\ \mathrm{cm}\sim 3\ R_{*}$, that is slightly higher than in classical SGXB ($a\approx 2\ R_{*}$).

The source goes through short episodes of highly-enhanced emission with flares reaching levels of a few $10^{36}\ \es$ as observed by \asca, \integral\ and \xmm. Only \swift\ did not detect any strong flare of the source during a short exposure time. \integral\ detected 6 flares in 4.8~Ms of effective exposure, so one flare every 9.2~days on average. This value can be considered as an upper limit as the average flare peak luminosity is just above the ISGRI sensitivity threshold when the source is within 9$\degr$ from the FOV centre. For a larger angular angle, fluxes of 4--5~\cps\ are below the 5$\sigma$ significance level (considering a pointing time-scale). As for half of the pointings (53\%) the source was farther than 9$\degr$ from the FOV centre, the bright flare frequency can be estimated as one every 4--5 days.

\subsection{Source spectra}

The spectra are fitted with models typically used for wind-fed accreting HMXB. The source is observed in  two states of luminosity, the quiescence phase and the flaring phase. Note that the \integral\ and \xmm\ observations are not simultaneous, explaining the high inter-calibration constant needed to fit the flare spectrum. For the quiescent spectrum that combines the average hard X-ray spectrum with an instantaneous X-ray spectrum, the inter-calibration constant is almost one. That implies that the accretion rate during the quiescent phase is rather constant. Therefore, the compact object would lie on a near-circular orbit and the stellar wind would be most of the time rather homogeneous ($\Delta\rho/\rho\sim2$).

The continuum is fitted with a flat power law $(\Gamma\sim 0.7-0.9)$ and has a high-energy cutoff $E_{\mathrm{cut}}\sim 16$~keV. The absorbing column density $\nh=(2.6\pm0.2)\times10^{22}\ \unit{cm}{-2}$ is higher than the whole interstellar absorption\footnote{The column density derived from the optical absorption $A_{\mathrm{V}}\sim7.6$ \citep{Coeal96} is $\nh\sim1.4\times10^{22}\ \unit{cm}{-2}$ using the relation derived by \citet{Predehlal95}.} expected on the line of sight $\nh\sim1.58\times10^{22}\ \unit{cm}{-2}$ \citep{Dickeyal90}. The observed additional absorption corresponds to the expected local absorption roughly estimated as $\nh\approx \dot{M}_{\mathrm{wind}}/(8\mu m_{\mathrm{H}}v_{\mathrm{wind}} a)\sim 2.4\times10^{22}\ \unit{cm}{-2}$ for a NS orbiting circularly at $a=3\ R_{*}$ around a massive star with spherically symmetric stellar wind.  When comparing the different observations of \asca, \swift\ and \xmm, the column density slightly varies by a factor 2 at maximum between $(1.6-3.6)\times 10^{22}\ \unit{cm}{-2}$ when the power law $\Gamma$ remains comparable in each case. Variation of the column density has been observed in other SGXB \citep[e.g.][]{Haberlal89,Pratal08} and explained by the interaction between the stellar wind and the X-ray emission of the compact object \citep{Blondinal90,Blondinal91}. For the few HMXB that have been monitored, these variations show higher factors such as 10. However, for \axj\, there are currently only 4 measures of $\nh$ and the orbital period of the system is not known.

A soft excess is also observed at low energies fitted with an optically-thin plasma with a temperature of $kT=0.18\pm0.05$~keV. Assuming $n_{\mathrm{e}}\sim n_{\mathrm{H}}\sim \dot{M}_{\mathrm{wind}}(a)/(4 \pi a^{2} \mu m_{\mathrm{H}} v_{\mathrm{wind}}(a))\sim10^{9}$~cm$^{-3}$ for $a\gtrsim 3\ R_{*}$, the size of the soft X-ray emitting region (sphere of radius $R_{\mathrm{em}}$) can be estimated from the mekal normalisation as $R_{\mathrm{em}}\approx \sqrt[3]{3\,N_{\mathrm{mekal}}/10^{-14}\times(D_{\mathrm{a}}/n_{\mathrm{H}})^{2}}$ resulting in $R_{\mathrm{em}}^{\mathrm{quies}}\sim10^{13}\ \mathrm{cm}\sim8R_{*}$ and $R_{\mathrm{em}}^{\mathrm{flare}}\sim3\times10^{13}\ \mathrm{cm}\sim20R_{*}$. Therefore, the soft X-ray excess is emitted within the stellar wind. Such excess has been observed in other HMXB and seems to be a common feature in these systems \citep{Hickoxal04}. They are likely due to X-ray scattering or partial ionization in the stellar wind \citep{Whiteal95}.

The second particular feature observed in the spectrum, specially in the quiescent one, is the absorption edge present at 7.9~keV. In other HMXB such as 4U~1700$-$37 \citep{Meeral05} or Vela~X$-$1 \citep{Nagaseal86}, absorption edges related to the K-edge absorption threshold of near-neutral iron have been observed with energies $E_{\mathrm{edge}}\gtrsim7.1$~keV. However, the edge energy detected in \axj\ is significantly higher and would correspond to highly-ionized iron of level XVIII and XIX \citep[see Fig.~7 in][]{Nagaseal86} or \citep[see Fig.~2 in][]{Kallmanal04}. The ionization parameter defined as $\xi=L_{\mathrm{X}}/n r_{\mathrm{X}}^{2}$ where $L_{\mathrm{X}}$ is the X-ray luminosity of the source, $n$ the gas density and $r_{\mathrm{X}}$ the distance from the X-ray source characterizes the photoionization of the gas \citep{Tarteral69}. For Fe XVIII and XIX, the limits of $\xi$ are $1<log(\xi)<3$ \citep[see Fig.~5 in][]{Kallmanal04}. Since $\tau=n \sigma_{\mathrm{Fe}} r_{\mathrm{X}}$ where $\sigma_{\mathrm{Fe}}$ is the iron cross-section \citep[that is $\sim10^{-20}-10^{-19}\ \unit{cm}{2}$ for Fe XVIII and XIX at 7.9~keV,][]{Kallmanal04}, the highly-ionized iron must be located at a distance within $1.5<log(L_{\mathrm{X}}\sigma_{\mathrm{Fe}}/\tau r_{\mathrm{X}})<2.5$, so $(6-63)\times10^{12}$~cm $\sim 4-40\,R_{\mathrm{*}}$, that is $\gtrsim a$, the expected binary separation as derived above. Thus, the highly-ionized iron is not located near the compact object but spread in the binary system.

\subsection{Flares by clumpy winds}

Most persistent SGXB show luminosity of the order of $10^{36}\ \es$ with frequent flares implying variation factors of the order of $\lesssim 20$. For \axj, we observe a persistent emission one order of magnitude lower than usual and seldom strong flares showing variation factors $\gtrsim 20$. Moreover, the accretion radius $R_{\mathrm{accr}}$ defined above implies a free-fall time of $t_{\mathrm{ff}}=0.5(\pi R^{3/2}/\sqrt{2GM})\sim 230$~s for matter accreted onto the compact object from $R_{\mathrm{accr}}\sim2\times 10^{10}$~cm. As the increase/decrease time of flares observed in \axj\ ($t_{\mathrm{incr/decr}}=100-850$~s) are also of this order, the infall matter producing the flares is thus radially accreted onto the compact object.

\citet{Walteral07} have shown that the flaring emission from transient SGXB, the so-called SFXT, must occur through accretion of clumpy winds onto a NS that is orbiting at a larger distance than usual ($\sim10\, R_{\mathrm{star}}$) and displaying very low quiescent emission (if observed) of $\lesssim10^{33} \es$ implying variability factors $\gtrsim 100$ \citep[see also][]{Negueruelaal08}.

In quiescence, \axj\ is more luminous than typical SFXT. It also features large variability with factors of 2--3 as observed in other SGXB. The source shows an intermediate behaviour between SFXT and SGXB. A clump/inter-clump density ratio of 45 is observed for \axj. Thus, the available $\dot{M}$ is lower than usual producing a persistent quiescent emission lower than in other SGXB but greater than in SFXT. The characteristics of the flares (number, luminosity, duration, increase/decrease time) observed in \axj\ imply that the source accretes some of these clumps ($M\sim10^{21-22}$~g) as explained in \citet{Walteral07}.

\section{Conclusion}\label{secCon}

\axj\ is most-likely a persistent X-ray binary with a O9.5I supergiant companion emitting at a rather low 0.2--100~keV luminosity of $\sim 10^{35}\ \es$ with seldom flares reaching luminosities of $10^{36}\ \es$. The most-accurate X-ray position is R.A.~(2000)~$=\ra{18}{45}{01.4}$ and Dec.~$=\dec{-04}{33}{57.7}$ (2$\arcsec$). Variability factors as high as 50 are observed on very short-time scale (few ks). The broad-band high-energy spectrum is typical of wind-fed accreting pulsars with a slight intrinsic absorption of $\nh=(2.6\pm0.2)\,10^{22}\ \unit{cm}{-2}$, a hard continuum of $\Gamma=(0.7-0.9)\pm0.1$, a high-energy cutoff at $\Ec=16_{-3}^{+5}$~keV. An excess at low energies is also observed fitted with a BB with a temperature of $kT=0.18\pm0.05$~keV. The presence of optically-thin and highly-ionised iron (Fe~XVIII$-$XIX) located far from the compact object but near the companion star ($\gtrsim 4\,R_{*}$) is also observed, principally during the quiescence phase of the source.  No spectral variations are observed between the flare and quiescence phases except the luminosity and the absorption depth of the 7.9~keV edge. The flare characteristics (number, luminosity, duration, increase/decrease time) in contrast to the persistent quiescent emission suggest that clumps of mass $M\sim 10^{22}$~g are formed within the stellar wind of the supergiant companion.

\begin{acknowledgements}
The authors  thank the anonymous referee his useful comments that improved the manuscript. The authors acknowledge the use of NASA's Astrophysics Data System. JAZH thanks J.~Rodriguez and S.~Chaty for giving useful comments.
\end{acknowledgements}

\bibliographystyle{aa}
\bibliography{../paper/biblio}

\end{document}